\documentclass[aps,pre, preprint, onecolumn]{revtex4}
\usepackage{amssymb}
\usepackage{eurosym}
\usepackage{epsfig}

\begin{document}

\normalsize

\title{Formation of a Vorticity Singularity in an Axisymmetric Poloidal Flow}
\author{E.\,A. Kuznetsov $^{1,2}$, E.\,A. Mikhailov $^{1,3}$, M.\,G.\,Serdukov $^3$}
\affiliation{$^{1}$ \textit{P.N. Lebedev Physical Institute of RAS, Moscow, Russia}\\
$^{2}$ \textit{L.D. Landau Theoretical Physics Institute of RAS, Chernogolovka, Russia}\\
$^{3}$ \textit{M.V. Lomonosov Moscow State University, Moscow, Russia}}

\begin{abstract}
In this work, building on previous studies of plane-parallel flows, we investigate the rapid growth of the frozen-in vorticity field in an axisymmetric poloidal flow. We present the results of numerical simulations performed using supercomputing resources. We show that exponentially growing velocity gradients can develop on the axis of symmetry, while vorticity gradients grow superexponentially. These results are qualitatively similar to those previously obtained for vorticity in a simpler two-dimensional flow.
\end{abstract}

\maketitle

\section{Introduction}

The problem of the formation of singularities in the flows of an inviscid incompressible fluid is one of the key problems of modern hydrodynamics. It is associated with the mechanism of energy transfer from large to small scales in fully developed turbulence. According to the Kolmogorov--Obukhov theory, for high values of Reynolds numbers ($Re \gg 1$) in the inertial range, velocity fluctuations at scale $\ell$ are of order $\langle \delta v\rangle \sim \ell^{1/3}$, while the corresponding vorticity fluctuations behave as $\langle \delta \omega\rangle \sim \ell^{-2/3}$ \cite{Kolmogorov1941,Obukhov1941}. Since the characteristic time of energy transfer remains finite, such a divergence at $\ell \to 0$ may indicate the possibility of the emergence of a vorticity singularity in a finite time, i.e. collapse within the framework of the Euler equation.

The possibility of collapse in the three-dimensional (3D) Euler equation has long been an open question. In 1984, Beale, Kato, and Majda \cite{BKM} showed that a necessary and sufficient condition for the existence of smooth solutions to the three-dimensional Euler equation describing ideal incompressible fluid flows over a finite time interval $[0, t_0]$ is the finiteness of the integral
\begin{equation} \label{BKM}
\int_0^{t_0}\mbox{sup}|{\bf \omega}| dt < \infty.
\end{equation}
Here, the supremum $|{\bf \omega}|$ is taken over the entire region $R^3$ under the assumption of finite velocity or under periodic boundary conditions. If the integral (\ref{BKM}) diverges, then the vorticity magnitude goes to infinity in a finite time. In other words, condition (\ref{BKM}) is a collapse criterion. The proof of this criterion is based on the theorem in \cite{Kato}.

Despite numerous attempts at numerical simulation,  it is still not entirely clean if the collapse is possible for Euler equations in three-dimensional case (see \cite{chae2008incompressible, gibbon2008three}). In two-dimensional case formation of singularities during finite time cannot be possible \cite{Wolibner1933,Kato1967,Yudovich1963}. Also, taking into account \cite{BKM}, Majda \cite{Majda1986} convincingly demonstated that for axisymmetric problems the results are nearly the same if there is now swirl (when the azimuthal velocity is zero) even if we take three-dimensinonal equations of ideal hydrodynamics.

However, nothing stays in the way of formation of such structures during infinite time periods. For two-dimensional equations, vorticity will have typical profiles with width decreasing according to exponential law, which has been shown using numerical modelling \cite{KuznetsovNaulin2007}. For three-dimensional case the computations showed that we can observe pancake-type structures that have thickness which becomes smaller exponentially \cite{Agafontsev2015,Agafontsev2016,Agafontsev2017}.

An important role in these processes is played by the property of vorticity being frozen into the fluid. For the vector field $\mathbf{\omega}$ frozen into a fluid, the following equation holds:
$$
    \partial_t \mathbf{\omega} = curl[\mathbf{u}\times\mathbf{\omega}], \qquad div \mathbf{u}=0.
$$
Although the flow itself is incompressible, frozen-in field lines can generally be compressed. This can lead to field concentration in small areas and to an increase in its gradients \cite{KuznetsovRuban1998,KuznetsovRuban2000,Kuznetsov2002}.

In the two-dimensional case, vorticity has a single non-zero component perpendicular to the flow plane. Therefore, it can be considered as a scalar quantity $\omega$. In such a flow, the growth of the vorticity itself is impossible, since its evolution is described by the transport equation
$$
    \partial_t \omega + (\mathbf{u},\nabla)\omega = 0, \qquad div \mathbf{u}=0,
$$
i.e., $\omega$ is a Lagrangian invariant. However, in this configuration, the curl of vorticity (known as ``divorticity'') is a frozen-in vector field,
$$
    \mathbf{B}=curl \mathbf{\omega},
$$
for which the equation holds
$$
    \partial_t \mathbf{B}
    =
   rot\,[\mathbf{u}\times\mathbf{B}],
    \qquad
   div\mathbf{u}=0.
$$
Therefore, due to the same compression mechanisms, the growth of the components of the curl of vorticity is possible, although the collapse of the vorticity itself is impossible in the two-dimensional case.

Note that in all the works discussed above, the question of the role of boundary conditions was not considered, in particular on a solid surface, when the normal velocity component vanishes under the condition of flow slipping along the boundary. Slip flows turn out to be compressible; they are characterized by an increase in velocity gradients due to the overturning effect, well known in gas dynamics. In this case, the compressibility of slip flows is accompanied by the appearance of jets in the direction normal to the boundary. As was shown in Ref. \cite{KiselevSverak2014}, in a two-dimensional flow in a disk described by the Euler equation, the velocity gradient grows according to an exponential law, reaching its maximum at the solid boundary, and the vorticity gradient there also grows according to a doubly exponential law. Moreover, in two-dimensional flows in regions with a non-smooth boundary, the formation of a singularity in a finite time at a finite kinetic energy \cite{KiselevZlatos2015} is possible.

In our previous studies \cite{KuznetsovMikhailov2022}, \cite{KuznetsovMikhailovSerdukov2023}, we showed that, for the inviscid Prandtl equation in both two ($d=2$) and three ($d=3$) dimensions, overturning of the slip flow at a solid boundary can lead to explosive growth of the velocity gradient for both $d=2$ and $d=3$ and, consequently, to the formation of a finite-time vorticity singularity for $d=3$. In the latter case, the three-dimensional collapsing solution has a tornado-like structure, with the singularity arising at the solid boundary. The explosive generation of vorticity results from the vortex drawing fluid from the near-surface layer. In Ref. \cite{KuznetsovMikhailovSerdukov2023}, we also numerically demonstrated, within the framework of the two-dimensional Euler equation, derivatives of velocity grow according to an exponential law, if the flows between to planes is studied. It has been shown that the gradient reaches its maximum at the boundary. As for vorticity gradient we have obtained double exponential growth.

In this paper we consider role of geometric factors for increasing of gradients of vorticity and velocity at the solid boundary of a cylinder in an axisymmetric poloidal incompressible flow in which vorticity is concentrated near the cylinder axis. Such flows are of considerable interest for applications, particularly in atmospheric physics.

In such a configuration, additional features arise due to the geometry, in particular due to the appearance of terms with $r^{-1}$ and $r^{-2}$ in the Laplacian. This requires specific approaches to numerical simulation, which leads to the use of significant computing resources.

We present numerical simulation results for the spatial and temporal evolution of vorticity, the vector potential of the velocity, and the velocity field. Superexponential growth of the velocity gradient is observed on the cylinder axis near the boundary, while the vorticity gradient exhibits behavior close to a double-exponential law. Numerical analysis predicts the formation of a rapidly contracting jet on the axis near the boundary within the toroidal vortex under study. This process may serve as a model for atmospheric phenomena such as the convective formation of cumulonimbus clouds and tracer transport.

\section{Governing Equations}

As noted in the Introduction, in the regime of fully developed turbulence, $Re \gg 1$, flow in the inertial range can be described in terms of the dimensionless three-dimensional Euler equation for an incompressible fluid
\begin{equation} \label{euler}
    \frac{\partial \mathbf{u}}{\partial t} + (\mathbf{u}, \nabla)\mathbf{u}
    = -grad p , \,\, div\mathbf{u} = 0.
\end{equation}
where $\mathbf{u} $ is the velocity, $p$ is the pressure, and the density of the medium $\rho$ is set equal to $1$.

Let us consider an axisymmetric flow, depending on the radial coordinate $r$, the coordinate $z$ along the cylinder axis and time $t$, inside a cylinder of radius $R$ and height $H$. In this work, we will limit ourselves to studying the dynamics of flows that have only two velocity components $u_r$ and $u_z$. It was for such flows (in the absence of azimuthal velocity, $u_{\phi}=0$) that Majda showed the absence of collapse \cite{Majda1986}, which became a natural generalization of the results of \cite{Wolibner1933,Kato1967,Yudovich1963} to axisymmetric flows. Note that the question of the influence of boundaries on the dynamics of vortices was not considered in Ref. \cite{Majda1986}.

For this axisymmetric flow, there are three equations for the two velocity components. The first is the incompressibility equation,
\begin{equation}
\frac{\partial u_z}{\partial z} + \frac 1r \frac{\partial} {\partial r}
ru_r=0,
\end{equation}
which is satisfied by introducing the stream function $\Psi$,
\begin{equation}
    u_r = -\frac{\partial \Psi}{\partial z},\quad u_z = \frac{1}{r}\frac{\partial}{\partial r}(r\Psi).
    \label{eq.1}
\end{equation}
The other two equations are:
\begin{equation} \label{Euler1}
\frac{d u_{r}}{d t}
        = -\frac{\partial p}{\partial r},\,\,
				\frac{d u_{z}}{d t}
        = -\frac{\partial p}{\partial z},
\end{equation}
where
\begin{equation}
  \frac{d}{d t} =  \frac{\partial}{\partial t} + u_r \frac{\partial}{\partial r }  + u_z\frac{\partial}{\partial z }.
\end{equation}
At $z=0, H$ and $r=R$, the cylinder boundary is subject to no-penetration conditions:
\begin{equation}
u_z(z = 0, H)  = 0,\,\,u_r(r = R) = 0.
\end{equation}
By symmetry, $u_r(r = 0) = 0$.

It is also easy to see that the vorticity $\mathbf{\omega}$ for such flows has only one non-zero azimuthal component:
$$
 \omega= \frac{\partial u_r}{\partial z} - \frac{\partial u_z}{\partial r},
$$
whence, using (\ref{eq.1}), we obtain the following expression for $\omega$ in terms of the stream function:
\begin{equation}
    \omega = -\frac{\partial}{\partial r}\frac{1}{r}\frac{\partial}{\partial r}(r \Psi) - \frac{\partial^2 \Psi}{\partial z^2}.
    \label{eq.2}
\end{equation}
The operator acting on $\Psi$ differs, up to sign, from the Laplacian by the additional term $r^{-2}$.

Next, in the equations (\ref{Euler1}), we eliminate the pressure and obtain equations for the vorticity $\omega$ and the stream function $\Psi$.

In an inviscid medium, vorticity is a frozen-in quantity and obeys the equation
$$
\frac{\partial \mathbf{\omega}}{\partial t} = rot[\mathbf{u}\times\mathbf{\omega}].
$$
In this case, for the azimuthal vorticity component $\omega$, this equation takes the form:
\begin{equation}
 \frac{\partial \omega}{\partial t} + u_r\frac{\partial \omega}{\partial r} + u_z\frac{\partial \omega}{\partial z} = \frac{1}{r}u_r\omega.
    \label{eq.3}
\end{equation}

Thus, the system describing the flow consists of the vorticity transport equation (\ref{eq.3}), the relation between $\omega$ and $\Psi$ given by (\ref{eq.2}), and the velocity expressions (\ref{eq.1}), supplemented by the initial and boundary conditions:
\begin{equation}
    \begin{cases}
        \partial_{t} \omega + u_r\frac{\partial \omega}{\partial r} + u_z\frac{\partial \omega}{\partial z} = \frac{1}{r}u_r\omega,\\
        \frac{\partial}{\partial r}(\frac{1}{r}\frac{\partial (r \Psi)}{\partial r}) + \frac{\partial^2 \Psi_\varphi}{\partial z^2} + \omega_\varphi = 0,\\
        \Psi|_{t = 0} = f(r, z),\\
        \Psi|_{r = 0} = \Psi|_{r = R} = \Psi|_{z = 0} = \Psi|_{z = H} = 0.
    \end{cases}
    \label{eq.4}
\end{equation}
The boundary conditions on $\Psi$ follow from the no-penetration conditions on the cylinder walls ($u_r|_{r=R} = 0$, $u_z|_{z=0} = u_z|_{z=H} =0$). At the same time, the axial symmetry condition $u_r|_{r=0} = 0$ is imposed.

Differentiating the first equation in (\ref{eq.4}) with respect to $z$ yields an equation for the $r$-component of the curl of vorticity, $B_r = -\frac{\partial \omega}{\partial z}$:
$$
\frac{d B_r}{d t} = -\frac{\partial u_r}{\partial z} \frac{\partial \omega_\varphi}{\partial r} - \frac{\partial u_z}{\partial z} B_r + \frac{1}{r}u_rB_r + \frac{1}{r}\frac{\partial u_r}{\partial z}\omega.
$$

On the axis of symmetry, the radial velocity component and its derivative with respect to \(z\) vanish: $u_r=0,\, \frac{\partial u_r}{\partial z}=0$. In addition, at the points \((r,z)=(0,0)\) and \((r,z)=(0,H)\), the azimuthal vorticity component $\omega$ and the $z$-component of velocity are also zero: $\omega=0$, $u_z=0$. Therefore, at these points, the equation for \(B_r\) simplifies to
$$
\frac{\partial B_r}{\partial t} = - B_r(\frac{\partial u_z}{\partial z} - \frac{1}{r}u_r).
$$
Note that if the radial velocity component behaves as $u_r\sim r$ in a neighborhood of $r=0$, then the following relation holds on the axis as a consequence of the vanishing velocity divergence:
$$
    \frac{1}{r}u_r=\frac{\partial u_r}{\partial r}=-\frac{1}{2}\frac{\partial u_z}{\partial z}.
$$
If in some neighborhood of $r=0$ the radial velocity component behaves like $u_r\sim r^l$ ($l>1$), then
$$
    \frac{1}{r}u_r=\frac{\partial u_r}{\partial r}=\frac{\partial u_z}{\partial z}=0.
$$
The following equality therefore holds:
$$
    \frac{\partial u_z}{\partial z} - \frac{1}{r}u_r = \frac{3}{2}\frac{\partial u_z}{\partial z}.
$$
As a result, at points \((r,z)=(0,0)\) and \((r,z)=(0,H)\) for $B_r$ we get:
$$
\frac{\partial B_r}{\partial t} = - \frac{3}{2}B_r\frac{\partial u_z}{\partial z}
$$
or
\begin{equation}
    B_r(t) = B_r(0)\exp(-\frac{3}{2}\int^t_0\frac{\partial u_z}{\partial z}dt).
    \label{eq:5}
\end{equation}

We also consider the advection of a passive tracer with concentration $q$. The tracer is assumed to have no effect on the fluid dynamics and to be transported by the velocity field $\mathbf{u}=u_r\mathbf{e_r}+u_z\mathbf{e_z}$. Including dissipation, the tracer transport equation in cylindrical coordinates takes the form \cite{Kotov2011}
\begin{equation}
\frac{d q}{d t} = \frac{1}{Re}\left( \frac{1}{r}\frac{\partial}{\partial r}\left(r\frac{\partial q}{\partial r}\right) + \frac{\partial^2 q}{\partial z^2}\right),
\label{eq.6}
\end{equation}
where
$$
\frac{d q}{d t} = \frac{\partial q}{\partial t} + u_r\frac{\partial q}{\partial r} + u_z\frac{\partial q}{\partial z}.
$$
Conservation of tracer mass and the no-penetration condition for the fluid at the boundary imply Neumann boundary conditions for $q$ on the cylinder surface:
\begin{equation}
\frac{\partial q}{\partial z}|_{z=0}=0,\qquad
\frac{\partial q}{\partial z}|_{z=H}=0,\qquad
\frac{\partial q}{\partial r}|_{r=R}=0.
\label{eq.7}
\end{equation}
By axial symmetry, the following condition holds on the cylinder axis $r=0$:
\begin{equation}
\frac{\partial q}{\partial r}|_{r=0}=0.
\label{eq.8}
\end{equation}
The initial tracer distribution is set by a function depending on $r$ and $z$
\begin{equation}
q|_{t=0}=q_0(r,z).
\label{eq.9}
\end{equation}

Thus, after determining the velocity field from the system (\ref{eq.4}), the transport of the tracer is described by the system of equations (\ref{eq.6})--(\ref{eq.9}).

\section{Numerical Solution}

The system (\ref{eq.4}), (\ref{eq.6})--(\ref{eq.9}) was solved numerically using finite-difference methods on a uniform grid. Consider one step of the algorithm. At the current time level, the velocity components $u_r$ and $u_z$ are first calculated using (\ref{eq.1}). An implicit finite-difference scheme for (\ref{eq.3}) is then used to determine the azimuthal vorticity component $\omega$ at boundary and interior points. The elliptic equation (\ref{eq.2}) is solved by relaxation using an evolutionarily factorized Rosenbrock scheme with parameter $1/2$ and a logarithmically spaced set of steps. Finally, the parabolic equation (\ref{eq.6}) is advanced by one step using the same evolutionarily factorized Rosenbrock scheme with parameter $1/2$ \cite{Kalitkin2013}. Accuracy was monitored through conservation of kinetic energy. The calculations were performed on a grid with $10^8$ spatial nodes, with energy conserved to six significant figures.

The large gradients arising in this problem require fine grids. To improve computational efficiency, the algorithm was parallelized using MPI and run on the Lomonosov-2 supercomputer.

\begin{figure}
    \centering
    \includegraphics[width=\columnwidth]{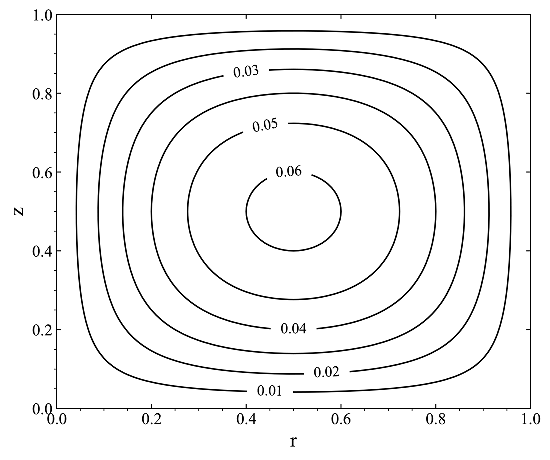}
    \caption{Contours of the azimuthal component of the vector potential $\Psi_\varphi$ at time $t=0$.}
    \label{Pic1}
\end{figure}

\begin{figure}
    \centering
    \includegraphics[width=\columnwidth]{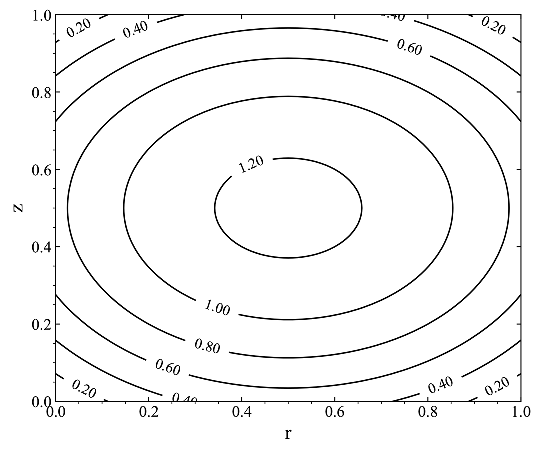}
    \caption{Contours of the azimuthal vorticity component $\omega$ at time $t=0$.}
    \label{Pic3}
\end{figure}

\begin{figure}
    \centering
    \includegraphics[width=\columnwidth]{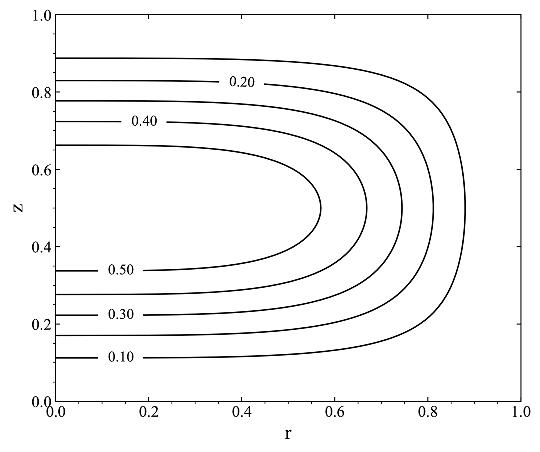}
    \caption{Contours of the concentration $q$ at time $t=0$.}
    \label{Pic5}
\end{figure}

Initially, the flow is a vortex rotating clockwise in the $(r,z)$ plane (Figs. \ref{Pic1} and \ref{Pic3}):
$$
    \Psi_\varphi|_{t=0}=r(r-R)z(z-H),\quad R = 1,\quad H = 1.
$$
The initial distribution of the concentration of a passive tracer was set as follows (Fig. \ref{Pic5}):
$$
    q|_{t=0}=(1-r^4)^2z^2(1-z)^2, \quad Re = 1000.
$$

\begin{figure}
    \centering
    \includegraphics[width=\columnwidth]{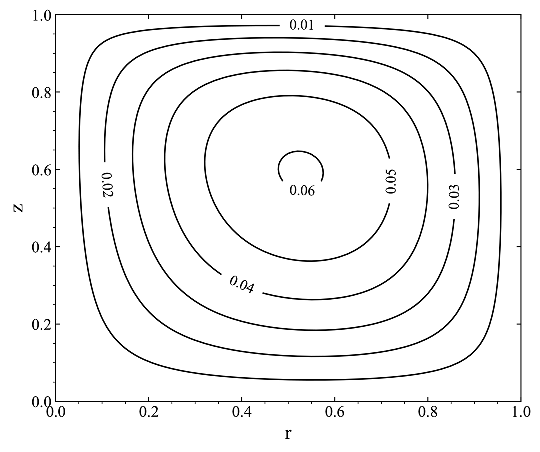}
    \caption{Contours of the stream function $\Psi$ at time $t=1.2$.}
    \label{Pic2}
\end{figure}

\begin{figure}
    \centering
    \includegraphics[width=\columnwidth]{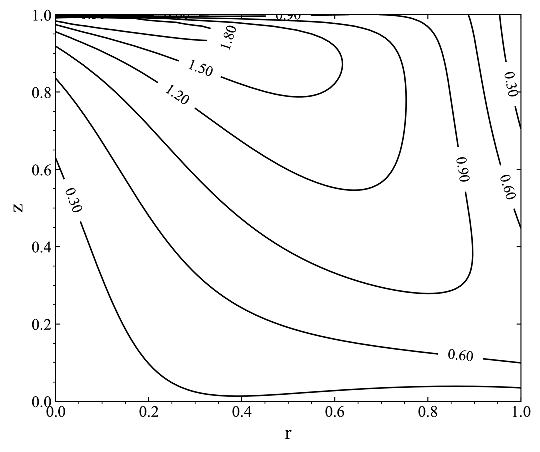}
    \caption{Contours of $\omega$ at time $t=1.2$.}
    \label{Pic4}
\end{figure}

\begin{figure}
    \centering
    \includegraphics[width=\columnwidth]{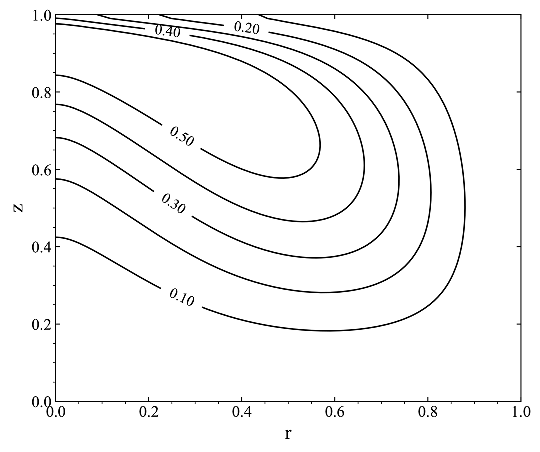}
    \caption{Contours of the concentration $q$ at time $t=1.2$.}
    \label{Pic6}
\end{figure}

\begin{figure}
    \centering
    \includegraphics[width=\columnwidth]{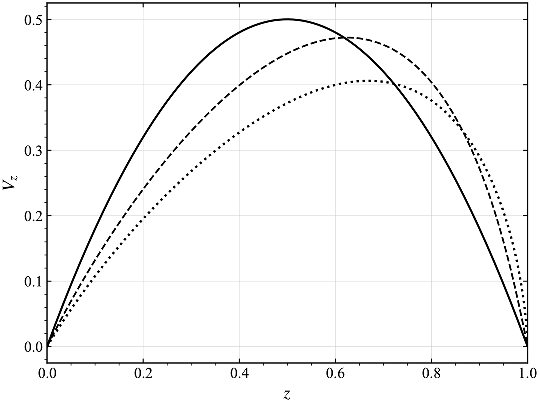}
    \caption{The $z$-component of velocity on the cylinder axis at different times: $t = 0$ (solid line), $t = 0.6$ (dashed line), and $t = 1.2$ (dotted line).}
    \label{Pic7}
\end{figure}

\begin{figure}
    \centering
    \includegraphics[width=\columnwidth]{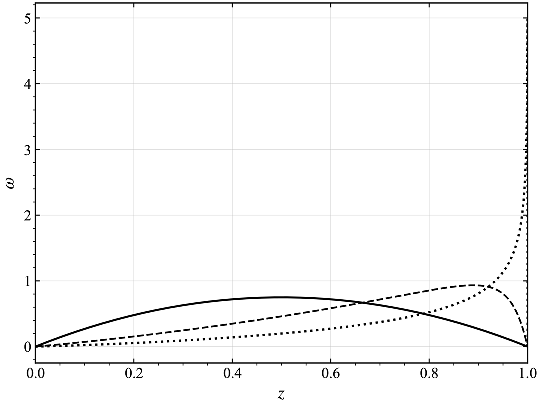}
    \caption{The azimuthal vorticity component on the cylinder axis at different times: $t = 0$ (solid line), $t = 0.6$ (dashed line), and $t = 1.2$ (dotted line).}
    \label{Pic8}
\end{figure}

\begin{figure}
    \centering
    \includegraphics[width=\columnwidth]{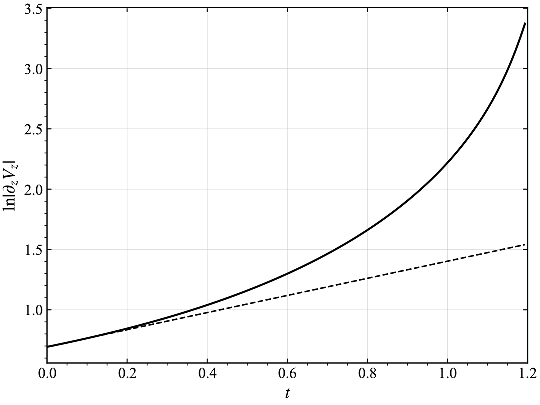}
    \caption{The derivative of the $z$-component of velocity with respect to $z$ at $r = 0$, $z = 1$ as a function of time on a logarithmic scale: numerical result (solid line) and exponential fit to the early-time behavior (dashed line).}
    \label{Pic9}
\end{figure}

\begin{figure}
    \centering
    \includegraphics[width=\columnwidth]{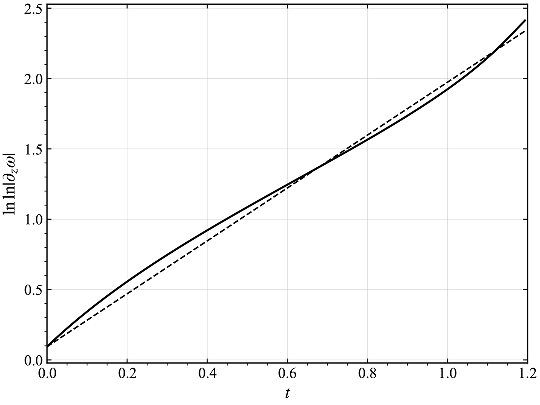}
    \caption{The derivative of the azimuthal vorticity component with respect to $z$ at $r = 0$, $z = 1$ as a function of time on a double-logarithmic scale: numerical result (solid line) and double-exponential fit (dashed line).}
    \label{Pic10}
\end{figure}

The numerical simulation results show that the flow (Figs. \ref{Pic2} and \ref{Pic4}), together with the tracer (Fig. \ref{Pic6}), begins to concentrate on the cylinder axis near the upper boundary. A process resembling overturning is observed near the upper end of the cylinder (Fig. \ref{Pic7}). As shown in Fig. \ref{Pic9}, the derivative of the $z$-component of velocity grows superexponentially and increases by a factor of 14.5 over the course of the evolution. At the same time, strong compression of the characteristic lines is observed for the azimuthal vorticity component $\omega$ (Fig. \ref{Pic8}). Figure \ref{Pic10} shows the evolution of the derivative of the azimuthal vorticity component. It grows according to a nearly double-exponential law and increases by a factor of more than $2\cdot10^4$ over the course of the evolution.

We also performed calculations with analogous initial conditions of the opposite sign, so that the vortex rotated counterclockwise. In this case, the flow concentrated on the axis near the lower boundary. Owing to the symmetry of the initial conditions, however, the absolute values of the derivatives at $(r, z)=(0,0)$ agree with those shown in Figs. \ref{Pic9} and \ref{Pic10} to 13 digits.

\section{Conclusion}

We have investigated inviscid poloidal flow without azimuthal rotation in a cylinder with rigid boundaries. Numerical simulations were used to obtain the evolution of vorticity, the vector potential of the velocity, and the velocity field. Superexponential growth of the velocity gradient is observed on the cylinder axis near the boundary, while the vorticity gradient exhibits behavior close to a double-exponential law. Numerical analysis predicts the formation of a rapidly contracting jet on the axis near the boundary within the toroidal vortex under study. We have also shown that a passive tracer in this flow concentrates on the cylinder axis near the boundary, in the same region where the rapidly contracting jet forms.

Thus, despite the geometric factor, numerical simulations of an axisymmetric poloidal flow confirm the main conclusion of our previous work on the influence of a solid plane boundary on the exponential growth of velocity gradients and the superexponential growth of the vorticity gradient in the two-dimensional Euler equation \cite{KuznetsovMikhailovSerdukov2023}. This work represents an intermediate step toward the more complex problem of the formation of tornado-like vortex structures. Two approaches to numerical simulation are being considered for the next stage. The first is to study vortex flows in $r-z$ geometry while including the azimuthal velocity component. The second, which more closely represents tornado-like vortex structures, is to solve numerically for three-dimensional flow between the planes $z=0$ and $z=h$, with periodic boundary conditions in the transverse directions. This problem could be solved using a fast Fourier sine transform in $z$, with the no-penetration condition at $z=0$ and $z=h$ ($v_z=0$), and a fast Fourier transform in the transverse coordinates.

The main result of this work is confirmation of the concept that singularities can form in finite or infinite time through the overturning of slip flow along a solid boundary. This motivates the study of more ambitious problems. In conclusion, we note that the numerical experiments of Luo and Hou on axisymmetric flow inside a cylinder of constant radius, with periodic conditions in $z$ along the cylinder axis, revealed explosive growth of vorticity at the solid boundary \cite{LuoHou2014}. Despite several mathematical studies \cite{ElgindiJeong1,ElgindiJeong2}, the central question remains whether collapse in this experiment results from overturning of the slip flow. The authors of \cite{ElgindiJeong1,ElgindiJeong2} considered flow outside the cylinder and showed that a vorticity singularity arises at the cylinder boundary in that case. A recent study~\cite{Cordoba} generalized these results and constructed an asymptotic solution for collapse in three-dimensional ideal hydrodynamics.
The use of neural-network approaches is also becoming increasingly widespread. For example, Ref.~\cite{Wang2023} demonstrated their applicability to solving the three-dimensional Euler equations and the two-dimensional Boussinesq equations. An initial approximation to the solution is constructed as a smooth function satisfying the conditions of the problem. It is then refined using artificial intelligence methods, yielding a solution that exhibits explosive growth in both cases.

This work was carried out using the equipment of the Center for Collective Use of High-Performance Computing Resources at M.V. Lomonosov Moscow State University.

\end{document}